\documentclass[twocolumn]{emulateapj}
\usepackage{ amsmath, color, natbib}
\newcommand{\kms}{km~s$^{-1}$}

\begin{document}
\title{A revised planetary nebula luminosity function distance to NGC 628 using MUSE}

\author{K. Kreckel\altaffilmark{1},  B. Groves\altaffilmark{2}, F. Bigiel\altaffilmark{3}, G. Blanc\altaffilmark{4,5}, J.~M.~D.~Kruijssen\altaffilmark{6}, A. Hughes\altaffilmark{7,8}, A. Schruba\altaffilmark{9}, E. Schinnerer\altaffilmark{1}}

\altaffiltext{1}{Max Planck Institut f\"{u}r Astronomie, K\"{o}nigstuhl 17, 69117 Heidelberg, Germany;  kreckel@mpia.de}
\altaffiltext{2}{Research School of Astronomy and Astrophysics, Australian National University, Canberra, ACT 2611, Australia}
\altaffiltext{3}{Institut f\"ur theoretische Astrophysik, Zentrum f\"ur Astronomie der Universit\"at Heidelberg, Albert-Ueberle Str. 2, 69120 Heidelberg, Germany}
\altaffiltext{4}{Departamento de Astronom?a, Universidad de Chile, Camino del Observatorio 1515, Las Condes, Santiago, Chile}
\altaffiltext{5}{Centro de Astrof?sica y Tecnolog?as Afines (CATA), Camino del Observatorio 1515, Las Condes, Santiago, Chile}
\altaffiltext{6}{Astronomisches Rechen-Institut, Zentrum f\"{u}r Astronomie der Universit\"{a}t Heidelberg, M\"{o}nchhofstra\ss e 12-14, 69120 Heidelberg, Germany}
\altaffiltext{7}{CNRS, IRAP, 9 Av. du Colonel Roche, BP 44346, F-31028 Toulouse cedex 4, France}
\altaffiltext{8}{Universit\'{e} de Toulouse, UPS-OMP, IRAP, F-31028 Toulouse cedex 4, France}
\altaffiltext{9}{Max-Planck-Institut f\"ur extraterrestrische Physik, Giessenbachstrasse 1, 85748 Garching, Germany}

\begin{abstract}

Distance uncertainties plague our understanding of the physical scales relevant to the physics of star formation in extragalactic studies.  The planetary nebulae luminosity function (PNLF) is one of very few techniques that can provide distance estimates to within $\sim$10\%, however it requires a planetary nebula (PN) sample that is uncontaminated by other ionizing sources.  We employ optical IFU spectroscopy using MUSE on the VLT to measure [OIII] line fluxes for  sources unresolved on 50 pc scales within the central star-forming galaxy disk of NGC 628.  We use diagnostic line ratios to identify 62 PNe,  30 supernova remnants and 87 HII regions within our fields.  Using the 36 brightest PNe we determine a new PNLF distance modulus of 29.91$^{+0.08}_{-0.13}$ mag (9.59$^{+0.35}_{-0.57}$ Mpc), in good agreement with literature values but significantly larger  than the previously reported PNLF distance.  We are able to explain the discrepancy and recover the previous result when we reintroduce SNR contaminants to our sample.  This demonstrates the power of full spectral information over narrowband imaging in isolating PNe. Given our limited spatial coverage within the galaxy, we show that this technique can be used to refine distance estimates even when  IFU observations cover only a fraction of a galaxy disk.

\end{abstract}

\section{Introduction}

When connecting star formation and galaxy evolution, nearby galaxies (D$<$20 Mpc) benefit from an external view while still accessible at high physical resolution (10-100pc) using ground based instruments ($\sim$1\arcsec seeing).  Yet a major source of uncertainty in the study of nearby galaxies is their distances.  Reliable methods for distance determination ($<$5\% accuracy)  based on Cepheids and tip of the red giant branch (TRGB) measurements are challenging to observe at large (D$>$5 Mpc) distances, requiring the resolution of individual stars \citep{Jacoby1992}.   Type II supernovae \citep{Dessart2005, Kasen2009} and Tully-Fisher \citep{Freedman2010} methods, often used in nearby galaxies, suffer from much larger ($\sim$20\%) uncertainties.  The [OIII] $\lambda$5007 \AA~ planetary nebula luminosity function (PNLF) provides an alternate, yet still accessible, method for obtaining accurate ($<$10 \%) distances \citep{Ciardullo1989, Jacoby1992, Ciardullo2010, Ciardullo2013}.  

Planetary nebulae (PNe) are  ionized by low and intermediate mass stars (1-8 M$_\sun$) with ages 0.1 - 1 Gyr as they evolve from the asymptotic giant branch to the white dwarf phase \citep{Paczynski1971, Iben1983}.  The central stars are very bright ($>$6000 L$_\sun$; \citealt{Vassiliadis1994}), with $\sim$12\% of the total luminosity of the central star reprocessed into the [OIII] $\lambda$5007~\AA~ emission line \citep{Dopita1992, Schonberner2007, Schonberner2010}, making them relatively easy to observe in emission line surveys.  The PNLF exhibits a sharp exponential cutoff, making the top $\sim$1 mag of the luminosity function well suited for use as an standard candle \citep{Ciardullo1989}.  Although theoretical models predict dependences on the stellar population age and metallicity \citep{Dopita1992, Marigo2004}, the PNLF has been empirically shown to be invariant across a variety of galaxy types (elliptical, spiral and irregular) that host very different underlying stellar populations \citep{Ciardullo1989, Feldmeier1996, Feldmeier1997}.  

Previous PNLF studies employed narrowband imaging, which is strongly contaminated by background stellar continuum emission from the galaxy. As a result, most of the identified PNe in these surveys are located at large radii \citep{Herrmann2008}, where the contamination is reduced but the PN density (which traces the stellar surface density) is lower \citep{Ciardullo1989}. Optical integral field spectroscopy provides full spectral information, allowing us to cleanly fit and remove emission from the stellar continuum.  In star forming spiral galaxies, this technique also allows us to model and remove background line emission arising from the diffuse interstellar medium \citep{Kreckel2016}, and provides simultaneous observation of a suite of bright emission lines (H$\alpha$, [NII] $\lambda$6584 \AA, [SII] $\lambda$6717 \AA, [SII] $\lambda$6737 \AA) that can be used  to distinguish PNe from other [OIII] emitters such as HII regions and supernova remnants \citep{Sabbadin1977, Baldwin1981, Riesgo2006}.  Past IFU studies have demonstrated the feasibility of this integral field unit (IFU) technique when extracting PN line fluxes \citep{Roth2004, Sarzi2011}, but lacked a large field of view and observed only a handful of PNe.  Recent and ongoing IFU surveys of nearby galaxies are typically designed to trace star formation and measure stellar populations, but in many cases will have sufficient depth and spatial resolution to simultaneously detect a large population of PNe, providing significant improvements on the distance estimates for these galaxies.  

We examine two regions within the nearby grand design spiral galaxy  NGC 628 using VLT/MUSE optical IFU spectroscopy that was obtained in order to study the contrast between spiral arm and interarm star formation \citep{Kreckel2016}.  Its face-on orientation ($i$=9$^\circ$; \citealt{Blanc2013}) minimizes cross-contamination of PNe with other objects and reduces the impact of internal dust extinction on our observations.  NGC 628 has a well studied metallicity gradient of 12 + log(O/H) = 8.834 - 0.485 $\times$ R (dex R$_{25}^{-1}$), as measured using ÒdirectÓ abundances based on observations of the temperature-sensitive auroral lines \citep{Berg2015}, with roughly solar abundances \citep{Asplund2009} within the central (R $<$ R$_{25}$) region.   
This galaxy has an uncertain distance, with estimates ranging from 7-10 Mpc based on a variety of techniques (see Section \ref{sec:comp}), including a narrowband PNLF study \citep{Herrmann2008}.  

In Section \ref{sec:data} we present our MUSE data. In Section \ref{sec:results} we identify PN candidates and, using diagnostic line ratios, exclude HII region and SNR contaminants from the sample.  In Section \ref{sec:discussion} we calculate a new PNLF distance to NGC 628, compare with previous distance estimates, explain our discrepancy with the previous PNLF distance, and explore what the limiting distance is for our technique.  We conclude in Section \ref{sec:conclusion}.  

\section{Data}
\label{sec:data}

We observed NGC 628 using the Multi-Unit Spectroscopic Explorer (MUSE; \citealt{Bacon2010}) at the Very
Large Telescope (VLT). This powerful new optical IFU provides a 1\arcmin $\times$ 1$\arcmin$ field of view with 0\farcs2 pixels and a typical spectral resolution of $\sim$2.75\AA(150 km s$^{-1}$).  We observed NGC 628 in two northern (ID 094.C-0623) and one southern (ID 095.C-0473) position,  with all data reduction details provided in \cite{Kreckel2016}.  Total on-source exposure times are 42 (50) minutes for our northern (southern) pointings. Observations were taken using the nominal wavelength range, covering 4800-9300\AA.  Typical seeing is 0\farcs8 across all fields.  Given the range of distances reported for this galaxy this corresponds to a spatial resolution of  30-50 pc.

\begin{figure}
\includegraphics[width=3.2in]{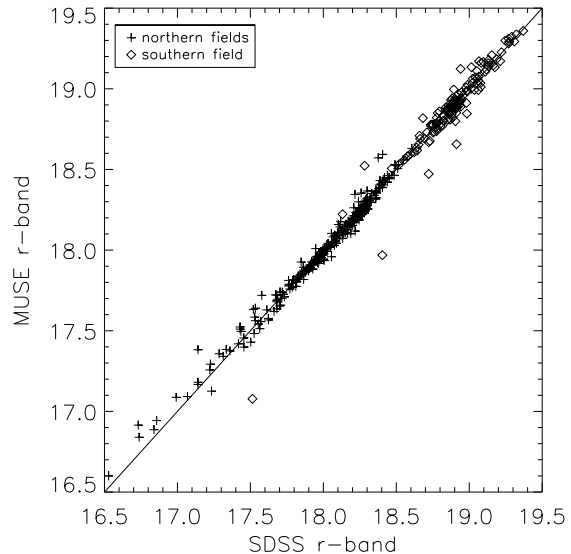}
\caption{Comparison of MUSE with SDSS photometry in the r-band for 5$\arcsec$ apertures across all fields.  Fluxes systematically agree to within 1\% (solid line shows a one-to-one agreement) with a scatter of less than 5\% (0.05 mag).  Outliers in the southern field lie at the edge of bright star clusters where we are sensitive to slight ($<$0\farcs2) offsets in astrometry. 
\label{fig:sdsscomp}}
\end{figure}

We validate the astrometry and flux calibration of the resulting MUSE data cubes against SDSS r-band images \citep{sdss}.  We construct a simulated r-band image from our data cube by convolving the r-band filter shape with the MUSE spectra at each spatial position.  We then fit compact sources in both the simulated MUSE r-band image and the SDSS r-band image using SExtractor \citep{Bertin1996}.  We find the source positions agree within 0\farcs2 across all fields.  To check the absolute flux calibration, we sample both images with randomly placed 5$\arcsec$ apertures, and find systematic agreement (Figure \ref{fig:sdsscomp}) to within 1\% and a scatter of less than 5\% (0.05 mag), confirming the high level of accuracy in our MUSE data set.  

\begin{figure*}
\centering
\includegraphics[width=5in]{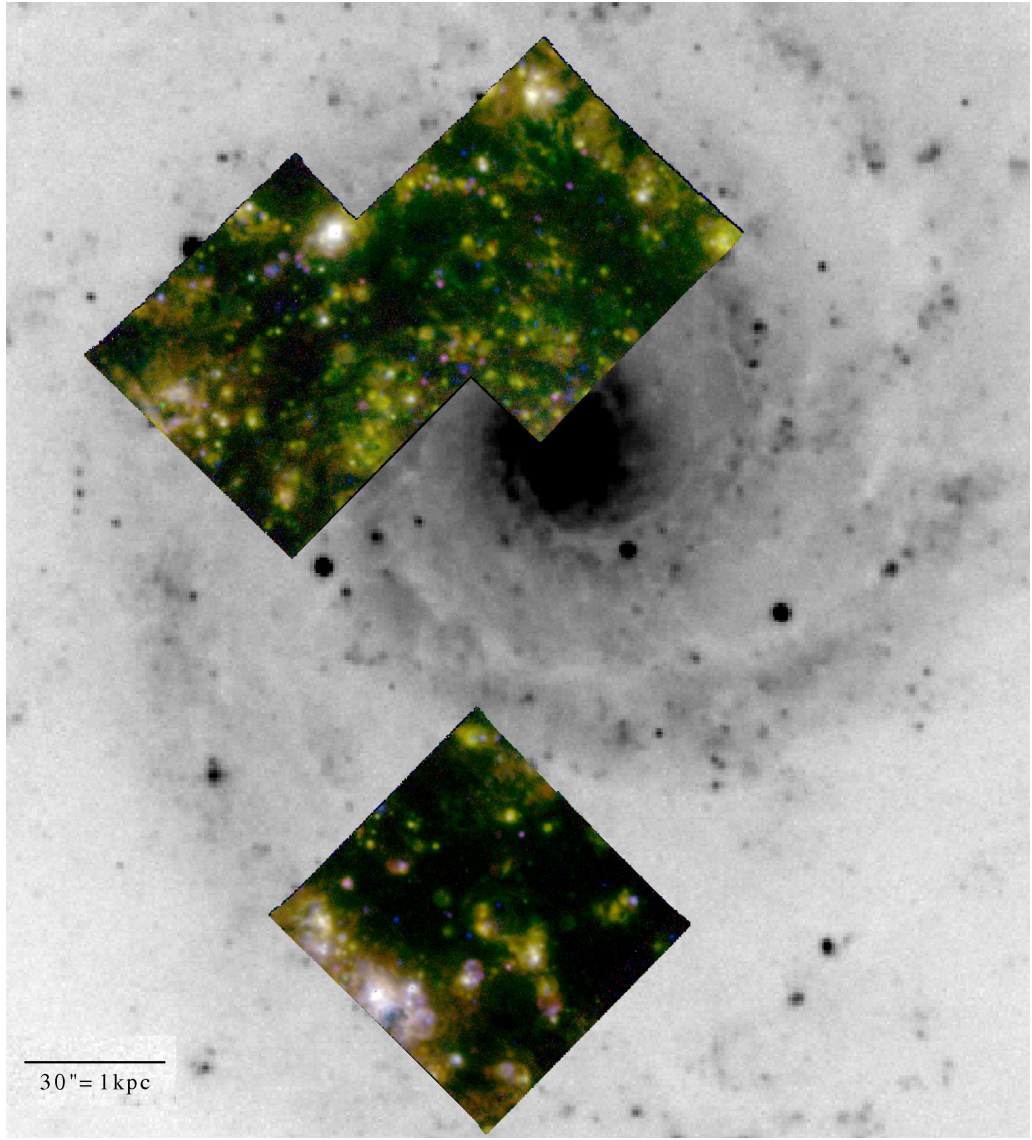}
\caption{V-band image of NGC 628 overlaid with three color images for the MUSE fields using three emission line maps: [OIII] (blue), H$\alpha$ (green) and [SII] (red).  These emission lines serve as diagnostics for the physical conditions in the ionized gas.  PN candidates appear as compact blue objects. 
\label{fig:img}}
\end{figure*}

Following \cite{Kreckel2016}, we fit the stellar continuum with MIUSCAT templates \citep{Vazdekis2012} and measure emission lines using LZIFU \citep{Ho2016} assuming single Gaussian fits.   These emission lines can serve as diagnostics for the mechanism that ionizes and excites the gas.   Figure \ref{fig:img} presents a three-color image of all fields, combining the [OIII], H$\alpha$ and [SII] line emission, highlighting the difference in color between the star-forming complexes and surrounding diffuse ionized gas as well as revealing many unresolved objects bright in [OIII], which are likely PNe (see Section \ref{sec:PNcandidates}).   

\begin{figure*}
\includegraphics[width=7in]{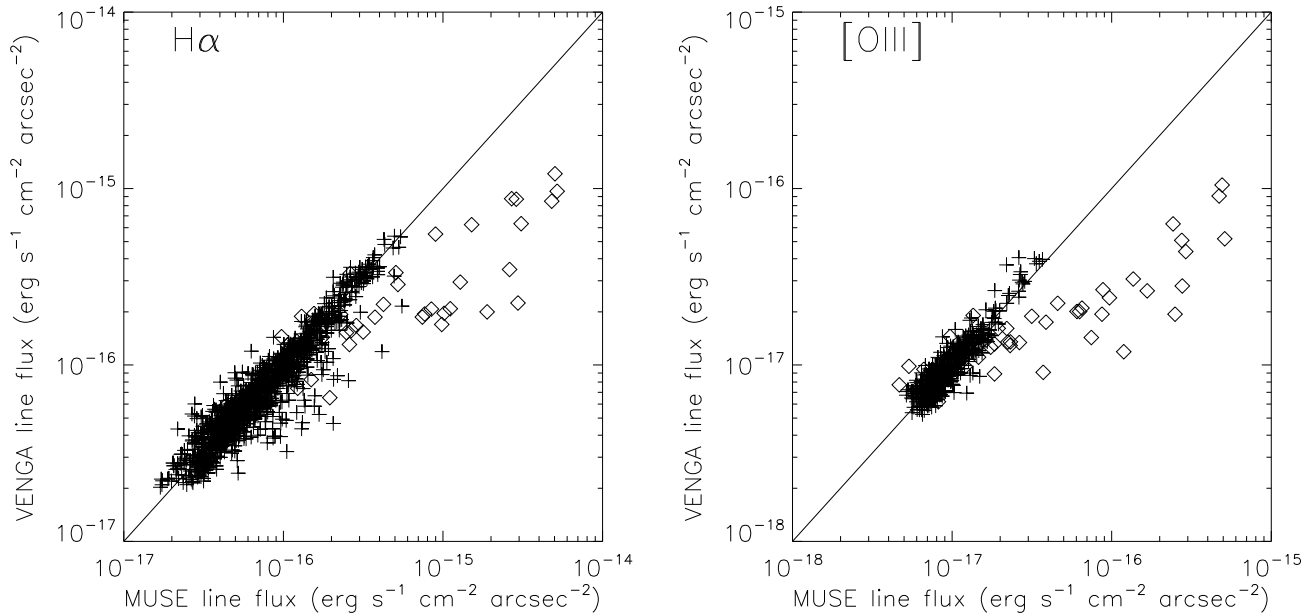}
\caption{Comparison of the H$\alpha$ and [OIII] line maps obtained in this study with those from VENGA \citep{Blanc2013}.  Line fluxes agree within 3\% (solid line shows a one-to-one agreement), with 20\% scatter (corresponding to 0.2 mag in the [OIII] line).   The large, order of magnitude offset observed in both lines towards high fluxes is at the position of an overlapping bright foreground star and HII region (diamonds), which also falls towards the edge of both the MUSE and VENGA field of view.  This combination of edge effects and extreme PSF wings invalidates our comparison near this region.  
\label{fig:vengacomp}}
\end{figure*}

As the PNLF is quite sensitive to the absolute flux calibration, we further check our measured line fluxes against the line fluxes observed by the optical IFU survey VENGA \citep{Blanc2013, Blanc2013a}.  
Here we have convolved our MUSE line maps to match the 5\farcs5 resolution in VENGA and compare the [O{\sc iii}] and H$\alpha$ line fluxes (Figure \ref{fig:vengacomp}).  We find our line fluxes agree within 3\%, with a scatter of 20\%.  For this comparison we exclude regions that show the largest discrepancy, as they are at the position of an overlapping bright HII region and a bright foreground star, which fall at the edge of the field of view in both images. The combination of edge effects and significant point spread function wings could bias our convolution, thus invalidating our comparison near this region.  
The overall agreement between the emission lines maps observed with these two very different instruments confirms that the systematics within the MUSE and VENGA data cubes are minor, and demonstrates the overall high quality of IFU observations possible in nearby galaxies. 

\section{Results}
\label{sec:results}

We identify our PN candidates based on unresolved objects in the [OIII] emission line maps.  However, we must also clean our candidate list of contamination by other line emitting objects, such as HII regions and supernovae remnants, before fitting the PNLF to determine a distance to the galaxy.

\subsection{Identifying PN Candidates}
\label{sec:PNcandidates}

\begin{figure*}
\includegraphics[height=2.8in]{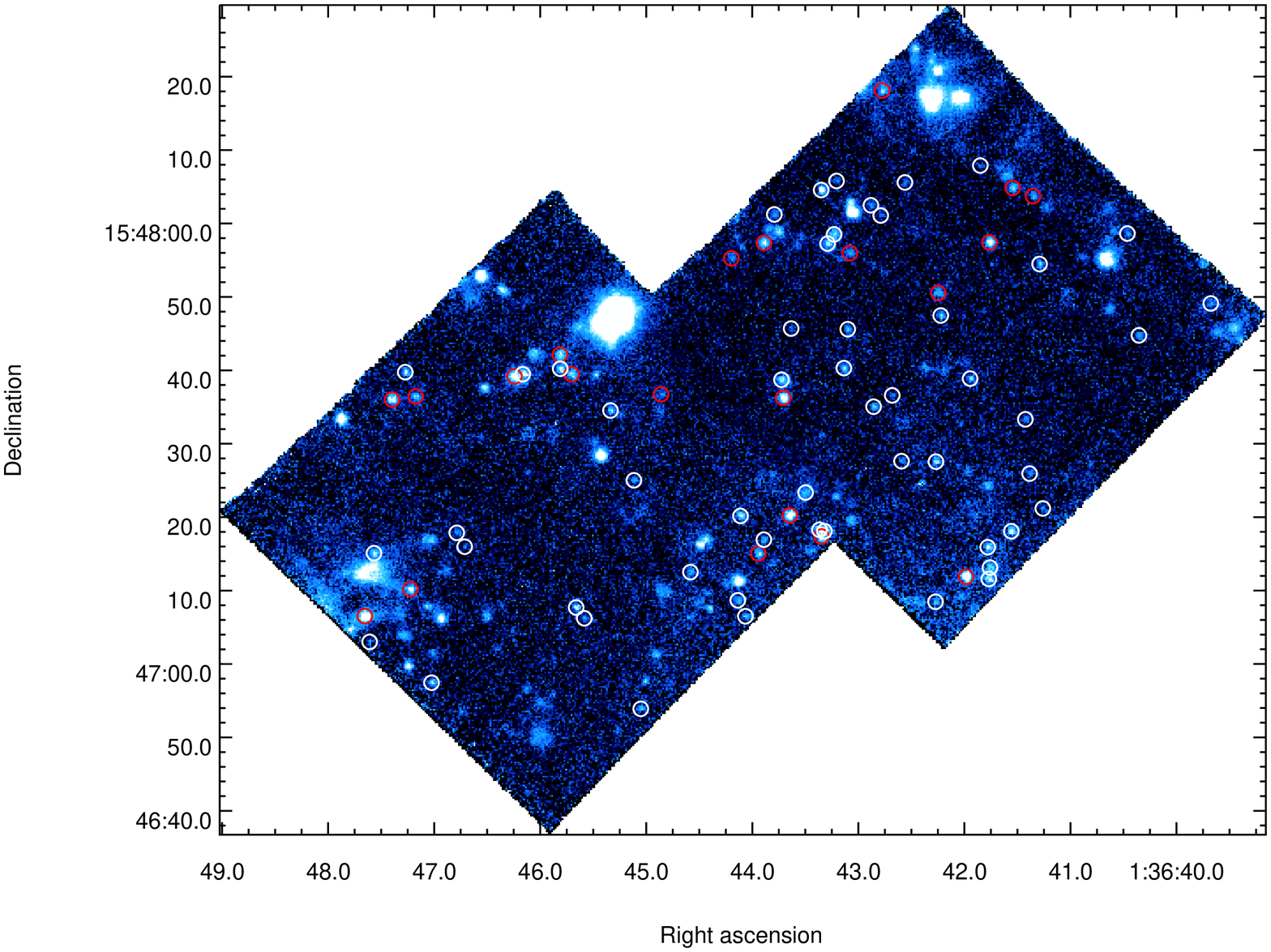}
\includegraphics[height=2.8in]{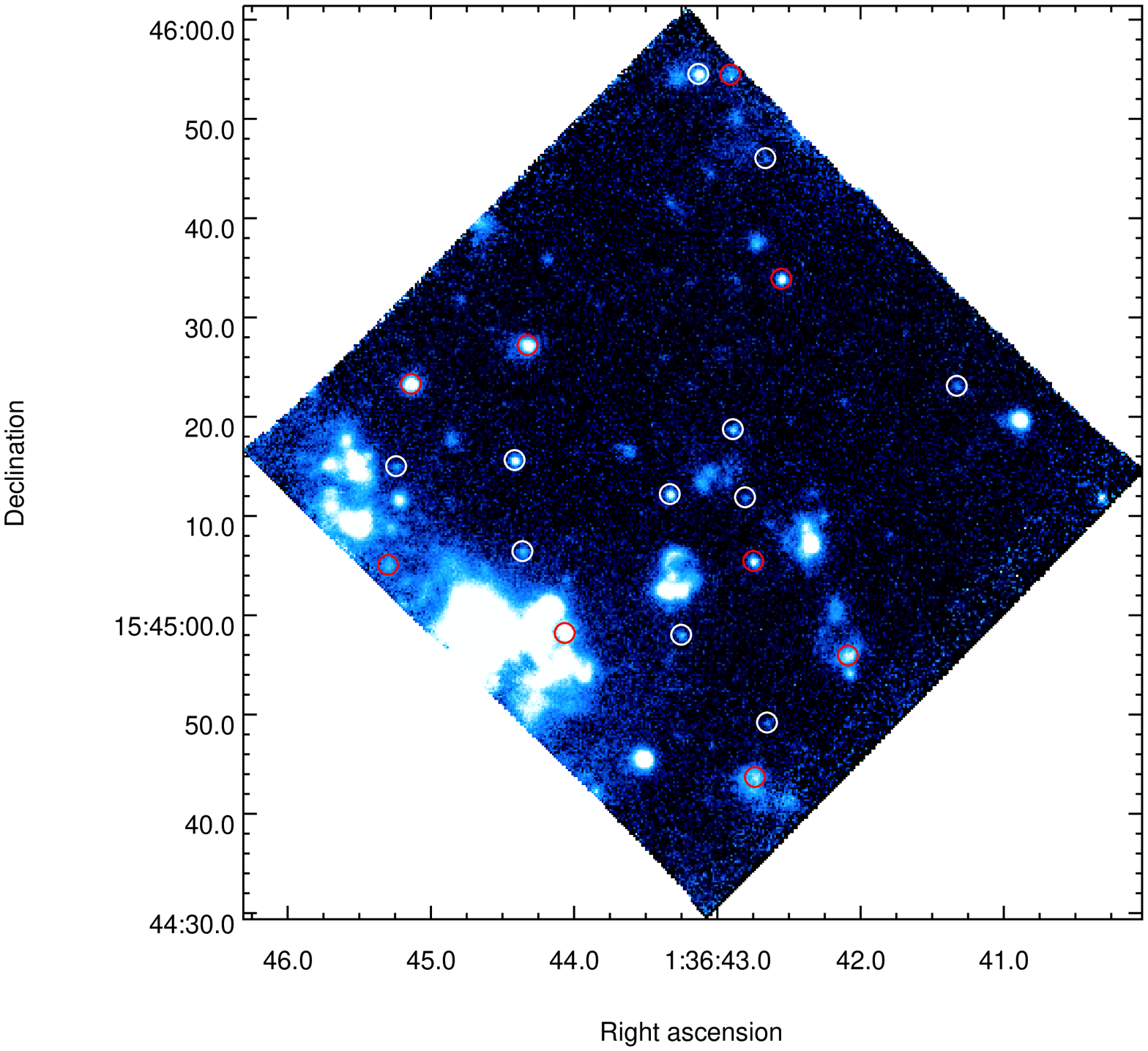}
\caption{[OIII] emission line maps for our MUSE fields.  Identified PNe (white circles) are uniformly distributed throughout the fields, consistent with their association with a moderately old (10$^8$-10$^9$ yr) stellar population. Compact SNRs (red circles) are a significant potential contaminant to our PN sample. 
\label{fig:imgoiii}}
\end{figure*}

As bright PNe in the Galaxy are much smaller than our 30-50 pc (0\farcs8) spatial resolution ( $<$ 1pc diameter; \citealt{Acker1992}), we use the IDL task FIND (an adaptation of DAOPHOT) to identify unresolved objects within the [OIII] line maps (Figure \ref{fig:imgoiii}).  We perform aperture photometry at each position for apertures  1\farcs6  (two times the FWHM)  in diameter.  As our field of view is entirely contained within the disk of the galaxy, we do not have any stars suitable for a standard curve of growth analysis to correct for lost flux outside of the aperture and robustly determine the seeing.  For this reason,  a slightly sub-optimal aperture size was chosen to conservatively ensure most of the flux is included in the aperture.  
In addition, as our line fluxes are measured from spectral fitting which does not allow for negative fluxes, our line maps are biased to positive fluxes in regions of low signal to noise.  This positive bias is not a background, and impacts only regions close to our detection limits.  Our apertures are small enough that this is not a concern when totaling the flux within the aperture, however it complicates any modeling of the aperture bias when attempting to recover flux from the wings of the PSF.  

There are no foreground stars within our field suitable for performing the aperture correction.  Therefore, we perform our curve of growth analysis on a model PSF, using parameters obtained from the brightest point source [OIII] line emitters.  This also avoids the bias to positive fluxes at regions with low signal to noise.  We convert these aperture corrected fluxes, F$_{[{\rm OIII}]}$, to an apparent magnitude as

\begin{equation}
m_{[{\rm OIII}]} = -2.5 \, \log \, F_{[{\rm OIII}]} - 13.74
\end{equation}
where F$_{[{\rm OIII}]}$ is given in ergs cm$^{-2}$s$^{-1}$ \citep{Jacoby1989}.

All objects are extinction corrected for the Milky Way foreground emission assuming  E(B-V)=0.062 \citep{Schlafly2011}, R$_V$=3.1 and the \cite{Cardelli1989} extinction law  (A$_V\sim$0.2 mag).    We apply no internal extinction correction based on the results of \cite{Feldmeier1997}, who argue that for face-on galaxies the expected extinction in m$_{[{\rm OIII}]}$ for face-on galaxies is less than 0.1 mag. This is supported by \cite{Herrmann2008}, who found that the PNLFs for the inner and outer disk of NGC 628 are the same.  As dust profiles are typically exponentially decreasing, with little extinction suffered in the outer disk \citep{Giovanelli1994, MunozMateos2009}, the lack of radial variation in the PNLF suggests that for NGC 628 there is also minimal extinction in central regions.     
Finally, from our stellar spectral fitting we observe reddening consistent with only the Milky Way foreground (A$_V$ $=$ 0.21 $\pm$0.15 mag), further supporting our decision not to apply any corrections for internal extinction.  

\begin{figure}
\includegraphics[width=3in]{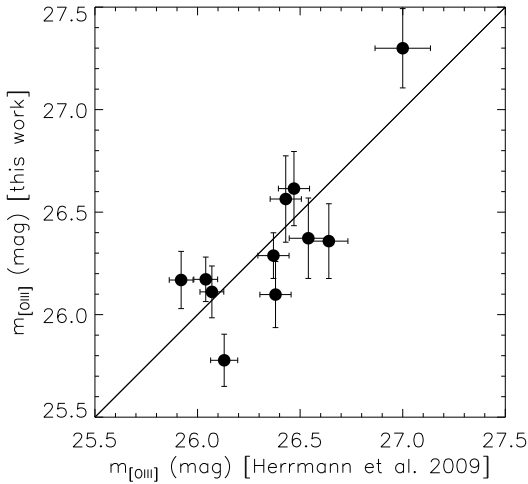}
\caption{A comparison of measured [OIII] apparent magnitude for 12 overlapping PNe observed in a previous narrowband imaging study \citep{Herrmann2008, Herrmann2009}.     We find very good ($\sim$1\%) systematic agreement between the two samples (solid line represents a one-to-one agreement), with most agreeing within 0.2 mag (20\%), close to our typical uncertainty. 
\label{fig:compHerrmann}}
\end{figure}

\cite{Herrmann2008} surveyed NGC 628 using [OIII] narrowband imaging and identified 153 PNe.  12 PNe are located within our field of view, and for these we compare m$_{[{\rm OIII}]}$ with their reported values (Figure \ref{fig:compHerrmann}).  We find reasonably good agreement between the two samples, with most agreeing within 0.2 mag, close to our typical uncertainty.  We find astrometric agreement within 1$\arcsec$ in all cases, with typical offsets of 0\farcs5. 

Based on our estimated uncertainties we are confident in our detections down to 3$\sigma$ = 27.0 mag in the northern fields and, due to the longer exposure time and slightly better observing conditions, 27.5 mag in the southern field. Due to the relatively low and uniform background (Figure \ref{fig:imgoiii}) we expect we are complete and robustly detect all sources down to these limits. 

\subsection{Removing contaminants}
\label{sec:contaminants}

Our [OIII] detected sample contains not just PNe but also compact HII regions or recent supernovae remnants (SNRs).  In order to identify contaminants to our PN sample, for all [OIII] detections we also measure line fluxes within the same aperture from H$\alpha$,  [NII] $\lambda$6584 \AA, [SII] $\lambda$6717 \AA, [SII] $\lambda$6737 \AA.  Combination of these emission lines provides diagnostic ratios that allow us to identify the most likely ionizing source for each of our [OIII] detections. Background galaxies, which contaminate narrowband imaging surveys, are ruled out here by our simultaneous detection of multiple emission lines for all objects.

More exotic ionizing sources are also possible.  Symbiotic binaries would be detectable through a very red stellar continuum spectrum, which we do not see for any of our sources.  We find three objects (Table \ref{tab:ems}) with very broad Balmer emission lines ($>$1000 km/s), strong He I emission and the CaII triplet in emission. Balmer lines in some sources show P Cygni profiles, and all sources fall in or near large star forming complexes.   These features are all consistent with Wolf-Rayet stars \citep{Crowther2007}, and we exclude these objects from our analysis.

\begin{deluxetable}{l c c }[h!]
\tablecaption{Emission line stars 
\label{tab:ems}}
\tablehead{
\colhead{Name}	&	
\colhead{Ra} &	\colhead{Dec} \\
\colhead{}	&	
\colhead{(J2000)} &	\colhead{(J2000)} 
}
\startdata
1 &  01:36:47.85 & +15:47:10.1 \\
2 & 01:36:44.17 & +15:47:16.0 \\
3 & 01:36:45.16 & +15:45:02.9 
  \enddata
\end{deluxetable}

Our aperture photometry may be biased by the background line emission arising from the diffuse ionized medium that is unrelated to the compact [OIII] sources \citep{Madsen2006,Haffner2009,Kreckel2016}.  Detailed two dimensional modeling of the diffuse background is necessary to ensure its careful removal from the line fluxes associated with the compact source \citep{Roth2004}.  However, as we are mainly interested in using these lines to identify contaminants in our PN sample we employ a simple median smoothing over 2$\arcsec$ scales (roughly three times the FWHM) to identify and remove the local diffuse background emission at the position of each [OIII] detection.  Masking discrete sources does not significantly change our results.  This background subtraction is only applied in cases where the median smoothed background is detected with signal to noise ratio greater than 5, which applies to most of the [NII], H$\alpha$ and [SII] line maps, but only 4\% of the [OIII] maps. We identify three PN candidates in positions where there is a high background in [OIII] and the detection lies superimposed on a star forming region.  Due to the difficulties in accurately subtracting the background, and increased likelihood of these being contaminants due to their position, we omit theses three objects from our analysis.  As the PNe are expected to be uniformly distributed throughout the disk, exclusion of this 4\% of our field of view should not bias our luminosity function.

\subsubsection{HII regions}
\label{sec:hii}

\begin{figure}[t!]
\centering
\includegraphics[width=3.5in]{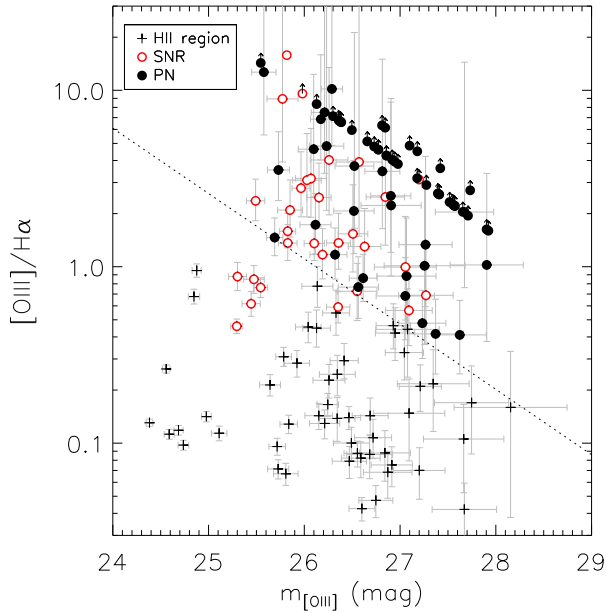}
\caption{[OIII]/H$\alpha$ line ratios as a function of [OIII] apparent magnitude for our full sample of [OIII] detected objects, including lower limits where H$\alpha$ is not detected at a 3$\sigma$ level.  1$\sigma$ uncertainties are shown.  We distinguish PNe (filled circles) from HII regions (plus symbols) following the diagnostic (dotted line) determined by \cite{Ciardullo2002} using local group PNe.   SNRs (open red circles) identified using additional line ratio diagnostics would significantly contaminate our PNe sample using only this threshold. All marked PNe form our final sample of PNe (Table \ref{tab:pns}). 
\label{fig:oiiiha}}
\end{figure}

For narrowband PN searches, H$\alpha$ imaging is commonly used to remove HII regions, which are the  main contaminant to our sample.  At the gas-phase metallicity of NGC\,628,  HII regions have H$\alpha$ flux brighter than [OIII] \citep{Shaver1983}, whereas the inverse is true in PNe \citep{Baldwin1981}.  \cite{Ciardullo2002} showed that PNe typically have line ratios of 

\begin{equation}
\label{eqn:oiiiha}
4 > \log{\frac{[OIII]}{H\alpha + [NII]}} > -0.37 M_{[{\rm OIII}]} - 1.16,
\end{equation}
where this has been empirically determined using the H$\alpha$+[NII] line due to contamination of [NII] within the narrowband filters.  
Other diagnostic diagrams are available, based on spectroscopic observations rather than narrowband imaging, to distinguish different ionizing sources \citep{Sabbadin1977, Baldwin1981, Riesgo2006}. However, due to the natural variations between PNe physical conditions and given that we have only  [OIII] and H$\alpha$ detected for many of our PNe, we find this empirical narrowband criteria the most effective.  
Assuming a 50\% contribution by [NII] \citep{Riesgo2006} and a distance modulus of 29.91 mag (see Section \ref{sec:dm}), we apply Equation \ref{eqn:oiiiha} to our sample to identify 87 HII region contaminants (Figure \ref{fig:oiiiha}).  Assuming a lower [NII] contribution \citep{Baldwin1981, Sabbadin1976} or varying the distance modulus within the uncertainty range does not significantly affect the resulting PNLF and distance estimate.  Many of our sources are not detected in H$\alpha$, and thus have only lower limits.  In addition, many sources appear to achieve ratios well above 4, however due to the large uncertainties in the H$\alpha$  line fluxes we retain these sources as potential PN candidates.

\subsubsection{Compact supernova remnants}
\label{sec:snr}

\begin{figure}[h!]
\centering
\includegraphics[width=3.2in]{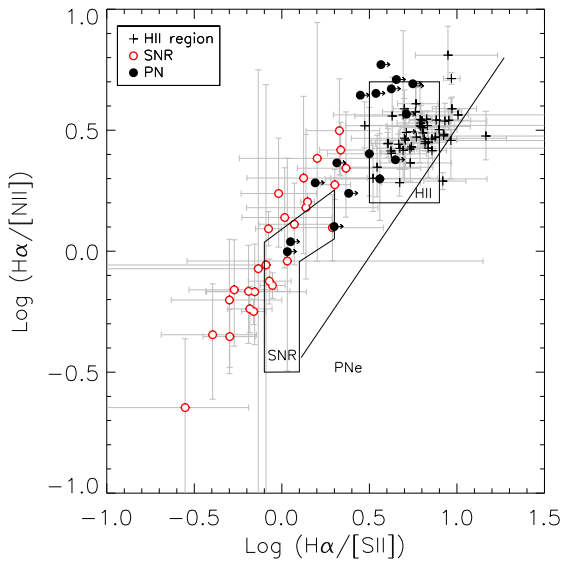}
\caption{H$\alpha$/[NII] as a function of H$\alpha$/[SII] for all sources with 3$\sigma$ detections in these emission lines.  1$\sigma$ uncertainties are shown.  The zones associated with HII regions, SNRs and PNe proposed by \cite{Sabbadin1977} are marked, however more recent work shows that PNe populate a broader range than this simple diagnostic line would suggest \citep{Riesgo2006}.  As described in the text, we identify SNRs (open red circles) and HII regions (plus symbols) by their position in the diagram. No PNe (filled circles) have detections in all three emission lines, and while those shown have lower limits on H$\alpha$/[SII] consistent with SNRs or HII regions, as they are not inconsistent with PNe we retain them in our PN sample.  All marked PNe are in our final sample of PNe (Table \ref{tab:pns})
\label{fig:siiha}}
\end{figure}

Another contaminant that is more difficult to identify and remove is compact supernova remnants (SNRs).  While the most evolved SNRs can span $\sim$100 pc, and are thus expected to be resolved by our survey, younger remnants are only $\sim$20 pc in size and will remain unresolved \citep{Franchetti2012}.  None of the three SN known within the galaxy (SN 2002ap, SN 2003gc, SN 2013ej) are within our observed fields, however previous study of diagnostic line ratios in SNR, HII region and PN populations \citep{Sabbadin1977, Riesgo2006} reveal that they inhabit very different regions of a diagram that compares H$\alpha$/[NII] and H$\alpha$/[SII] line ratios (Figure \ref{fig:siiha}).  Many of our PN candidates are not detected in any of these lines, however we can still use the line ratio when detected to identify contaminants to our sample.

We find that the  HII regions identified in Figure \ref{fig:oiiiha} are consistent with the expected region of the diagram, and find a population of sources with line ratios consistent with SNRs. 
Due to the difficulties in accurately subtracting the diffuse background emission from the emission lines other than [OIII] and the faintness of these lines,  the uncertainties on our line ratio measurements are large, and we employ a simple threshold to identify all sources with detected line ratios  H$\alpha$/[SII] $>$ 2.5 as SNRs \citep{Blair2004}.  A few of our PN candidates have line ratio detections or limits that place them in regions typically populated by HII regions or SNRs, however given  that these sources are not inconsistent with PNe (within the uncertainties) we retain them in our sample.    

We identify a total of 30 SNR candidates (Figure \ref{fig:imgoiii}; Table \ref{tab:snr}), including one object (\#17) previously identified as a PN (M74-30) by \cite{Herrmann2009}. Our fields do not cover the six SNRs previously identified from narrowband imaging \citep{Sonbas2010}. We do not expect this sample to necessarily be complete as we are identifying only those SNRs which are compact enough to be confused with PNe.  
As the SNRs with the highest [OIII] are expected to have strong shocks, we measure the velocity dispersion in the [OIII] line and find a median value of $\sim$200 \kms, slightly resolved given the instrumental dispersion of 150 \kms and further confirming our identification of these sources as SNRs.  This includes the SNR  \#17, previously identified as a PN, for which we measure a velocity dispersion  of $\sim$250 \kms.  
The median velocity dispersion for our PNe sample, with expected intrinsic line widths of $\sim$20 \kms \citep{Richer2010a, Richer2010b}, is consistent with being unresolved. Also due to the shock excitation, SNRs are expected to have strong [OI] compared to [OIII], in contrast to PNe where [OI] should be 100 times fainter \citep{Baldwin1981}.  Our data is not deep enough to place strong limits on the [OI] flux, however, we detect [OI] for 17 of our SNR and but only one PN (\#13, which may  indicate it has been misclassified).   
These SNRs significantly contaminate our sample, as determined by our [OIII]/H$\alpha$ diagnostic (Figure \ref{fig:oiiiha}).  This has significant implications for the derived PN luminosity function and measured distance (see Section \ref{sec:comp}).

\begin{deluxetable*}{l c c c c c c }[h!]
\tablecaption{SNR Identifications. 
\label{tab:snr}}
\tablehead{
\colhead{Name}	&	
\colhead{Ra} &	\colhead{Dec} &	 \colhead{m$_{[{\rm OIII}]}$} &	\colhead{[OIII]/H$\alpha$} &	\colhead{H$\alpha$/[SII]} &	\colhead{H$\alpha$/[NII]}\\
\colhead{}	&	
\colhead{(J2000)} &	\colhead{(J2000)} &	 \colhead{(mag)} &	\colhead{} &	\colhead{} &	\colhead{}
}
\startdata
   1 &  01:36:45.14  & +15:45:23.3 & 25.29 &  0.46 &  1.40 &  1.60 \\
  2 &  01:36:47.65  & +15:47:06.5 & 25.30 &  0.88 &  2.17 &  2.62 \\
  3 &  01:36:41.98  & +15:47:11.9 & 25.44 &  0.62 &  0.88 &  0.72 \\
  4 &  01:36:43.65  & +15:47:20.2 & 25.47 &  0.85 &  0.69 &  0.56 \\
  5 &  01:36:46.24  & +15:47:39.2 & 25.49 &  2.36 &  2.32 &  2.20 \\
  6 &  01:36:44.33  & +15:45:27.2 & 25.54 &  0.76 &  2.01 &  1.88 \\
  7 &  01:36:43.35  & +15:47:17.4 & 25.77 &  8.94 &  0.73 &  0.85 \\
  8 &  01:36:44.07  & +15:44:58.2 & 25.82 & $>$15.82 & $<$ 0.16 & - \\
  9 &  01:36:42.09  & +15:44:56.0 & 25.83 &  1.59 &  2.13 &  3.15 \\
 10 &  01:36:43.71  & +15:47:36.2 & 25.83 &  1.36 &  0.66 &  0.58 \\
 11 &  01:36:47.40  & +15:47:36.0 & 25.85 &  2.09 &  1.04 &  1.38 \\
 12 &  01:36:42.74  & +15:44:43.7 & 25.97 &  2.79 &  1.33 &  2.01 \\
 13 &  01:36:47.22  & +15:47:10.2 & 25.98 & $>$ 9.57 & $<$ 0.16 & - \\
 14 &  01:36:45.81  & +15:47:42.1 & 26.03 &  3.09 &  0.40 &  0.45 \\
 15 &  01:36:45.70  & +15:47:39.5 & 26.07 &  3.16 &  0.50 &  0.63 \\
 16 &  01:36:42.55  & +15:45:33.9 & 26.10 &  1.36 &  0.85 &  0.75 \\
 17\tablenotemark{a} &  01:36:41.76  & +15:47:57.4 & 26.15 &  2.47 &  0.70 &  0.68 \\
 18 &  01:36:43.89  & +15:47:57.4 & 26.19 &  1.17 &  0.50 &  0.44 \\
 19 &  01:36:43.94  & +15:47:15.1 & 26.26 &  4.03 &  0.28 &  0.23 \\
 20 &  01:36:42.75  & +15:45:05.5 & 26.35 &  0.59 &  0.84 &  1.24 \\
 21 &  01:36:45.30  & +15:45:05.1 & 26.36 &  1.36 &  0.96 &  1.73 \\
 22 &  01:36:42.77  & +15:48:18.1 & 26.51 &  1.54 &  1.59 &  2.42 \\
 23 &  01:36:47.17  & +15:47:36.5 & 26.55 &  0.73 &  1.94 &  1.25 \\
 24 &  01:36:42.91  & +15:45:54.5 & 26.57 &  3.92 &  0.81 &  0.88 \\
 25 &  01:36:41.54  & +15:48:04.8 & 26.63 &  1.30 &  0.65 &  0.68 \\
 26 &  01:36:42.25  & +15:47:50.5 & 26.85 &  2.48 &  1.07 &  0.91 \\
 27 &  01:36:44.86  & +15:47:36.7 & 27.06 &  1.00 &  0.53 &  0.69 \\
 28 &  01:36:43.08  & +15:47:56.0 & 27.09 &  0.57 &  1.18 &  1.29 \\
 29 &  01:36:41.35  & +15:48:03.7 & 27.20 & $>$ 3.12 & $<$ 0.63 & - \\
 30 &  01:36:44.20  & +15:47:55.3 & 27.27 &  0.69 &  1.37 &  1.51 
 
  \enddata
    \tablenotetext{a}{identified as a PN in \cite{Herrmann2009}}
\end{deluxetable*}

\subsection{Calculating the distance}
\label{sec:dm}

\begin{deluxetable*}{l c c c c c c}[h!]
\tablecaption{Planetary Nebula Identifications. 
\label{tab:pns}}
\tablehead{
\colhead{Name}	&	
\colhead{Ra} &	\colhead{Dec} &	 \colhead{m$_{[{\rm OIII}]}$} &	\colhead{[OIII]/H$\alpha$} &	\colhead{H$\alpha$/[SII]} &	\colhead{H$\alpha$/[NII]} \\
\colhead{}	&	
\colhead{(J2000)} &	\colhead{(J2000)} &	 \colhead{(mag)} &	\colhead{} &	\colhead{} &	\colhead{} 
}
\startdata 
 1 &  01:36:43.32  & +15:47:18.1 & 25.55 & $>$14.29 & - & - \\
  2\tablenotemark{a} &  01:36:43.37  & +15:47:18.3 & 25.58 & 12.67 & $>$ 0.77 & $>$ 1.35 \\
  3 &  01:36:43.13  & +15:45:54.5 & 25.69 &  1.46 &  3.16 &  2.53 \\
  4 &  01:36:46.16  & +15:47:39.5 & 25.73 &  3.53 & $>$ 2.41 &  1.73 \\
  5\tablenotemark{a} &  01:36:44.11  & +15:47:20.2 & 26.10 &  4.64 & $>$ 1.31 & $>$ 2.29 \\
  6 &  01:36:45.81  & +15:47:40.2 & 26.12 &  1.73 & $>$ 3.45 &  4.49 \\
  7 &  01:36:47.57  & +15:47:15.1 & 26.13 & $>$ 8.35 & - & - \\
  8\tablenotemark{a} &  01:36:44.42  & +15:45:15.6 & 26.17 &  6.86 & $>$ 1.27 & $>$ 2.16 \\
  9\tablenotemark{a} &  01:36:43.35  & +15:48:04.6 & 26.21 &  7.49 & $>$ 0.73 & $>$ 1.28 \\
 10 &  01:36:41.77  & +15:47:11.6 & 26.23 &  4.83 & $>$ 1.11 & $>$ 1.94 \\
 11\tablenotemark{a} &  01:36:43.33  & +15:45:12.2 & 26.29 & 10.18 & $>$ 0.77 & $>$ 1.31 \\
 12 &  01:36:41.56  & +15:47:18.1 & 26.30 & $>$ 7.13 & - & - \\
 13 &  01:36:43.29  & +15:47:57.3 & 26.32 &  1.17 & $>$ 4.22 &  4.69 \\
 14\tablenotemark{a} &  01:36:47.02  & +15:46:57.5 & 26.36 & $>$ 6.76 & - & - \\
 15\tablenotemark{a} &  01:36:47.27  & +15:47:39.7 & 26.37 & $>$ 6.67 & - & - \\
 16 &  01:36:43.72  & +15:47:38.7 & 26.38 & $>$ 6.62 & - & - \\
 17 &  01:36:43.23  & +15:47:58.5 & 26.39 & $>$ 6.59 & - & - \\
 18 &  01:36:41.76  & +15:47:13.1 & 26.50 & $>$ 5.95 & - & - \\
 19 &  01:36:45.66  & +15:47:07.7 & 26.52 &  2.07 & $>$ 1.99 &  1.26 \\
 20 &  01:36:41.78  & +15:47:15.9 & 26.52 &  3.72 & $>$ 1.10 & $>$ 1.93 \\
 21\tablenotemark{a} &  01:36:45.34  & +15:47:34.5 & 26.56 &  0.77 & $>$ 5.15 &  3.69 \\
 22 &  01:36:43.50  & +15:47:23.3 & 26.61 &  0.86 &  3.63 &  1.99 \\
 23\tablenotemark{a} &  01:36:43.14  & +15:47:40.3 & 26.66 & $>$ 5.13 & - & - \\
 24 &  01:36:42.27  & +15:47:27.6 & 26.73 & $>$ 4.80 & - & - \\
 25 &  01:36:44.14  & +15:47:08.7 & 26.77 & $>$ 4.61 & - & - \\
 26 &  01:36:45.05  & +15:46:53.9 & 26.77 & $>$ 4.61 & - & - \\
 27 &  01:36:44.36  & +15:45:06.5 & 26.81 & $>$ 6.33 & - & - \\
 28 &  01:36:42.22  & +15:47:47.5 & 26.81 &  3.47 & $>$ 0.90 & $>$ 1.58 \\
 29 &  01:36:42.89  & +15:45:18.7 & 26.84 & $>$ 6.14 & - & - \\
 30 &  01:36:43.89  & +15:47:16.9 & 26.86 & $>$ 4.27 & - & - \\
 31 &  01:36:44.58  & +15:47:12.5 & 26.86 & $>$ 4.26 & - & - \\
 32 &  01:36:41.95  & +15:47:38.9 & 26.90 &  2.52 & $>$ 1.15 & $>$ 2.01 \\
 33 &  01:36:44.07  & +15:47:06.5 & 26.91 &  2.23 & $>$ 1.29 & $>$ 2.27 \\
 34 &  01:36:45.12  & +15:47:25.0 & 26.92 & $>$ 4.04 & - & - \\
 35 &  01:36:41.29  & +15:47:54.5 & 26.94 & $>$ 3.94 & - & - \\
 36 &  01:36:42.86  & +15:47:35.0 & 26.98 & $>$ 3.82 & - & - \\
 37 &  01:36:46.79  & +15:47:17.9 & 27.05 &  0.68 & $>$ 3.68 &  5.90 \\
 38 &  01:36:42.27  & +15:47:08.5 & 27.07 &  0.88 & $>$ 2.81 &  4.41 \\
 39\tablenotemark{a} &  01:36:43.25  & +15:44:58.0 & 27.10 & $>$ 4.85 & - & - \\
 40 &  01:36:43.10  & +15:47:45.6 & 27.18 & $>$ 3.17 & - & - \\
 41 &  01:36:45.24  & +15:45:15.0 & 27.18 & $>$ 4.51 & - & - \\
 42 &  01:36:43.79  & +15:48:01.2 & 27.23 &  0.48 & $>$ 4.45 &  2.39 \\
 43 &  01:36:47.61  & +15:47:03.0 & 27.26 &  1.01 & $>$ 2.06 &  2.32 \\
 44 &  01:36:39.68  & +15:47:49.1 & 27.27 &  1.33 & $>$ 1.55 &  1.92 \\
 45 &  01:36:40.35  & +15:47:44.8 & 27.27 & $>$ 2.91 & - & - \\
 46 &  01:36:41.38  & +15:47:25.9 & 27.37 &  0.42 & $>$ 4.51 &  5.12 \\
 47 &  01:36:42.88  & +15:48:02.5 & 27.39 & $>$ 2.60 & - & - \\
 48 &  01:36:45.58  & +15:47:06.2 & 27.41 & $>$ 2.57 & - & - \\
 49 &  01:36:40.46  & +15:47:58.6 & 27.41 & $>$ 2.57 & - & - \\
 50 &  01:36:41.33  & +15:45:23.1 & 27.42 & $>$ 3.61 & - & - \\
 51 &  01:36:41.42  & +15:47:33.3 & 27.52 & $>$ 2.32 & - & - \\
 52 &  01:36:42.56  & +15:48:05.6 & 27.55 & $>$ 2.25 & - & - \\
 53 &  01:36:43.21  & +15:48:05.8 & 27.57 & $>$ 2.22 & - & - \\
 54 &  01:36:42.59  & +15:47:27.6 & 27.57 & $>$ 2.21 & - & - \\
 55 &  01:36:41.26  & +15:47:21.2 & 27.57 & $>$ 2.21 & - & - \\
 56 &  01:36:42.81  & +15:45:11.9 & 27.62 &  0.41 & $>$ 5.59 &  4.92 \\
 57 &  01:36:42.79  & +15:48:01.1 & 27.66 & $>$ 2.05 & - & - \\
 58 &  01:36:42.67  & +15:45:46.0 & 27.67 &  2.04 & $>$ 1.08 &  1.00 \\
 59 &  01:36:46.71  & +15:47:16.0 & 27.71 & $>$ 1.95 & - & - \\
 60 &  01:36:42.65  & +15:44:49.2 & 27.73 & $>$ 2.71 & - & - \\
 61 &  01:36:43.63  & +15:47:45.7 & 27.90 & $>$ 1.63 & - & - \\
 62 &  01:36:42.68  & +15:47:36.6 & 27.90 &  1.02 & $>$ 1.12 &  1.09 \\
 63 &  01:36:41.85  & +15:48:07.9 & 27.92 & $>$ 1.60 & - & - 

 \enddata
    
    \tablenotetext{a}{identified in \cite{Herrmann2009}}

\end{deluxetable*}

After applying our line diagnostic tests, we identify 63 PNe within our MUSE fields (Figure \ref{fig:imgoiii}; Table \ref{tab:pns}), including the 11 PNe previously identified by \cite{Herrmann2009} that fall within our field of view.  One additional PN from their catalog we reclassify as a SNR (see Section \ref{sec:snr})  

\begin{figure}[h]
\centering
\includegraphics[width=3.2in]{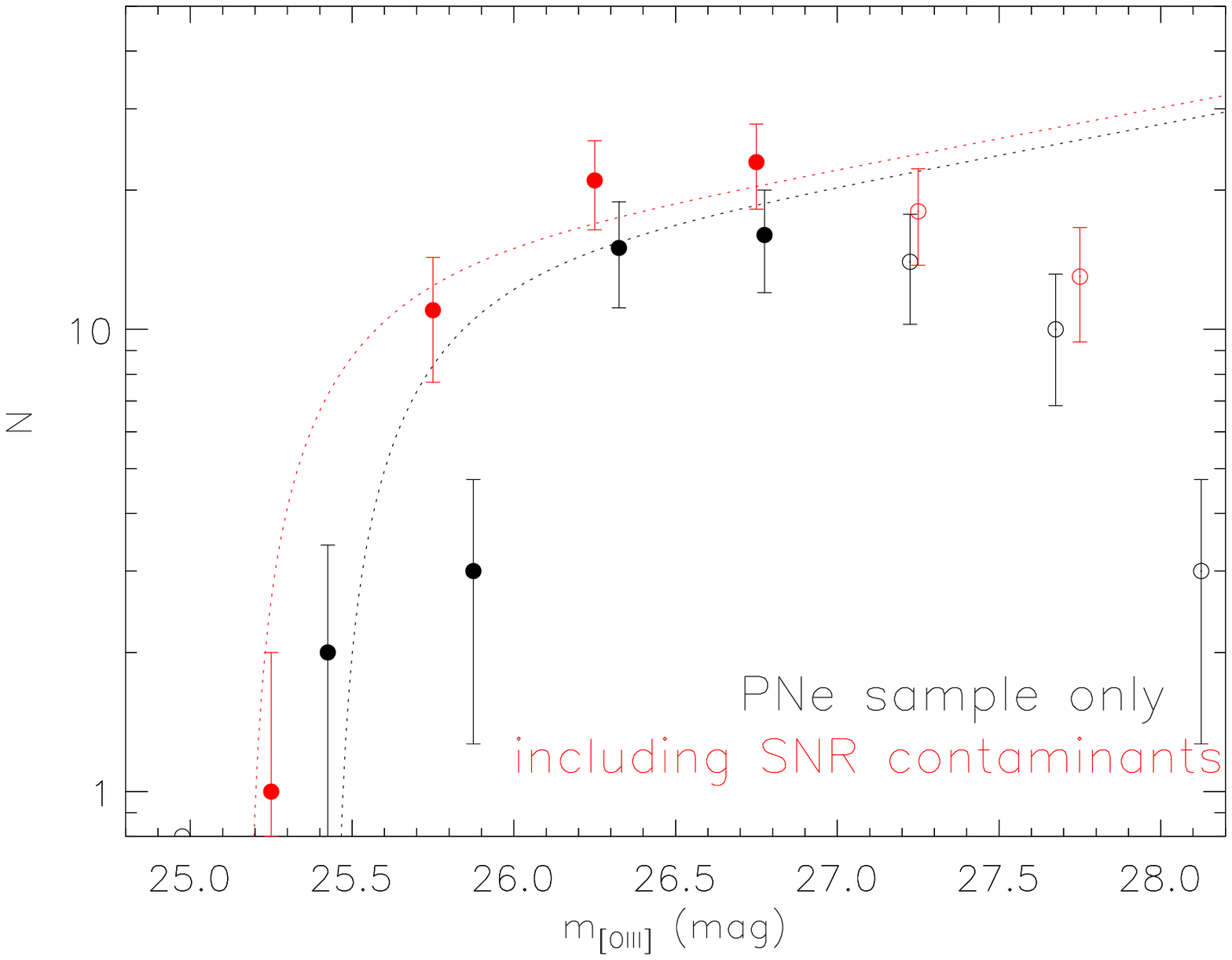}
\caption{Planetary Nebulae luminosity function using only our 63 identified PNe (black) and including contamination from 26 SNRs (red). Poisson error bars are included for each bin.  Our fits (dotted lines) are done using a maximum likelihood method on our sources brighter than the 27 mag completeness limit (filled symbols), allowing us to measure a distance modulus of 29.91$^{+0.08}_{-0.13}$ mag for NGC 628.    Introducing the SNR contaminants, we find a distance modulus of 29.65$^{+0.08}_{-0.13}$ mag, reproducing the previous PNLF estimate based on narrowband imaging \citep{Herrmann2008}
\label{fig:pnlf}}
\end{figure}

\begin{figure}
\centering
\includegraphics[width=3.5in]{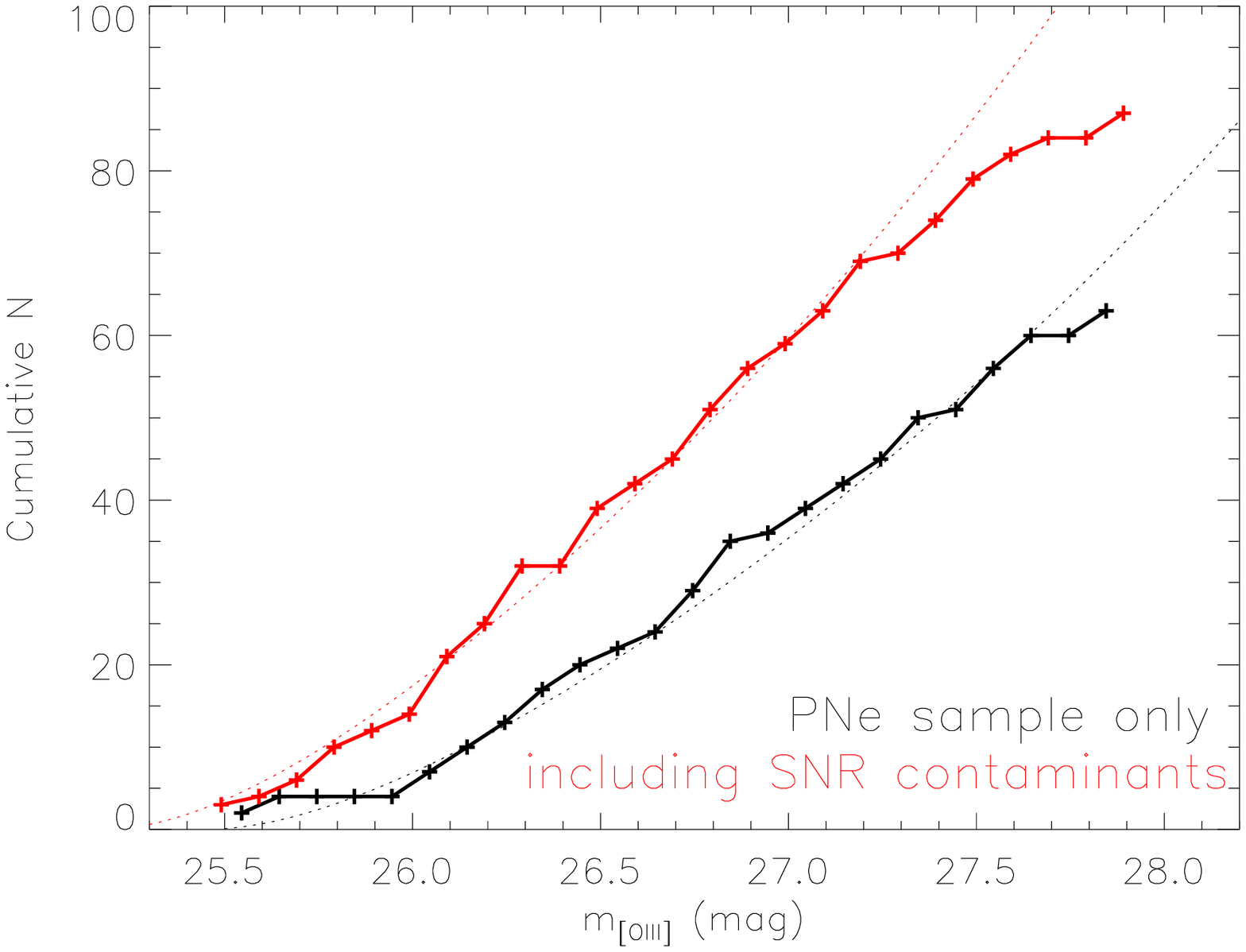}
\caption{Cumulative planetary Nebulae luminosity function for our 63 identified PNe (black) and including contamination from 26 SNRs (red).  While our low number statistics make it difficult to judge the quality of the PNLF fit  in Figure \ref{fig:pnlf}, we see here the good agreement between our fit (dotted lines) and the observed populations down to our completeness limit of 27 mag. 
\label{fig:cpnlf}}
\end{figure}

The planetary nebula luminosity function (PNLF) has been empirically shown to be well fit by an equation of the form

\begin{equation}
N(M) \propto e^{0.307M}\left( 1-e^{3(M^*-M)} \right)
\end{equation}
where M is the absolute magnitude and M$^*$ is the absolute magnitude of the most luminous PN \citep{Ciardullo1989, Jacoby1992}.  We adopt here the value of M$^*$ = -4.47$^{+0.02}_{0.03}$ mag, which applies for solar metallicity galaxies such as NGC 628 \citep{Ciardullo2002}.  Revised calibrations suggest a slightly higher value (M$^*$ = -4.53$\pm$0.06; \citealt{Ciardullo2012}), however this would change our distance by only $\sim$ 3\% and is consistent within the uncertainties.  
The PNLF cutoff is observed to be fainter in low-metallicity systems, consistent with theoretical models \citep{Dopita1992, Schonberner2010}.    
Using the method of maximum likelihood \citep{Ciardullo1989}, we determine the distance modulus that best fits the PNLF to our observations.  This method accounts for the decreased probability of observing PNe near the luminosity cutoff M$^*$, especially important given our small sample size, and removes any biases introduced by binning.  We include in our fit only the 36 PNe brighter than 27 mag (Figure \ref{fig:pnlf}), our expected completeness limit, and measure a distance modulus of 29.91$^{+0.08}_{-0.13}$ mag (9.59$^{+0.35}_{-0.57}$ Mpc).  Here our uncertainty includes (added in quadrature) the uncertainty in the fit (29.91$^{+0.06}_{-0.12}$ mag), our absolute photometric uncertainty ($\sim$0.03 mag), and a Monte Carlo sampling of our measured apparent magnitude ($\sim$0.05 mag).  Our low number statistics make it difficult to judge the quality of the PNLF fit, however the good agreement between our fit and observed PNe sample is apparent when examining the cumulative PNLF (Figure \ref{fig:cpnlf}; \citealt{Mendez1993}).  

\section{Discussion}
\label{sec:discussion}

Our revised PNLF distance is significantly larger than that previously determined using narrowband imaging, but consistent with distances obtained through other methods.  We test how the careful exclusion of SNRs affects our distance estimate, and show that this contamination can significantly bias our result.  As IFU observations are crucial to isolating a clean PN sample, we further explore the potential for using PNe as distance estimators in ongoing and future IFU studies.  

\subsection{Comparison with previous work}
\label{sec:comp}
A previous narrowband imaging survey of PNe in NGC 628 measured a distance modulus of 29.67$^{+0.06}_{-0.07}$ mag \citep{Herrmann2008}, significantly closer than our measurement.  In our comparison with their catalog (see Section \ref{sec:PNcandidates}), we identified 12 of their PNe that fall within our field of view and find no systematic offset between the measured [OIII] apparent magnitudes (Figure \ref{fig:compHerrmann}).  However, following our source classification using diagnostic line ratios (Section \ref{sec:contaminants}), we reclassify one of their PNe as a SNR.  Given that in Figure \ref{fig:oiiiha} we find that 26 SNRs would be classified as PNe without the additional emission line diagnostics provided by our IFU data,  we test how the introduction of these SNRs into our PNLF biases our resulting distance estimate.  If we fail to clean the SNR contaminants from our sample, our PNLF fit is still very good (Figure \ref{fig:pnlf}), and we measure a distance modulus of 29.65$^{+0.08}_{-0.13}$ mag, in very good agreement with \cite{Herrmann2008}.  

\begin{figure}
\centering
\includegraphics[width=3in]{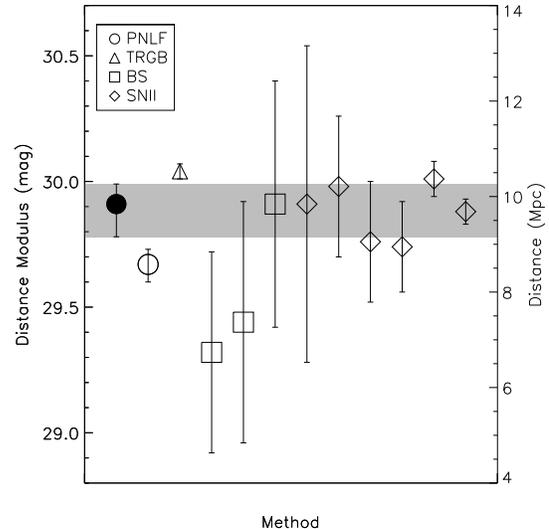}
\caption{Comparison of our distance estimate (filled circle) with values in the literature.  1$\sigma$ uncertainties are shown for each measurement.  Our revised PNLF distance is in much closer agreement (within 2$\sigma$) with the tip of the red giant branch (TRGB) measurement (triangle) than the previous PNLF distance (open circle).  We also find good agreement with most of the distances measured using the brightest supergiant (BS, squares) and SN II (diamonds) methods, though these all have much larger error bars.  See Table \ref{tab:dms} for detailed references for all methods.
\label{fig:lit}}
\end{figure}

We also compare our measured distance modulus to results from the literature using a variety of methods (Figure \ref{fig:lit}; Table \ref{tab:dms}).  We find good agreement with the most recent SN II methods, and 2$\sigma$ agreement with the recent TRGB measurement, which has much smaller uncertainties than all other methods.  

 \begin{deluxetable*}{c c c l }[h!]
\tablecaption{Comparison of Distance Moduli
\label{tab:dms}}
\tablehead{
\colhead{(m-M)}	&	
\colhead{D} &	\colhead{Method} &	\colhead{Reference} \\
\colhead{(mag)}	&	
\colhead{(Mpc)} &	\colhead{}  &	\colhead{}  
}
\startdata
29.32 $\pm$ 0.40 & 7.31 $\pm$ 1.35 & Brightest supergiants & \cite{Sharina1996}  \\
29.44 $\pm$ 0.48 & 7.73 $\pm$ 1.71 & Brightest supergiants & \cite{Hendry2005} \\
29.91 $\pm$0.49 & 9.59 $\pm$ 2.16 & Brightest supergiants (alternate calibration) & \cite{Hendry2005} \\
29.91$\pm$0.63 & 9.59 $\pm$ 2.78 & SN II standard candle  &  \cite{Hendry2005} \\
29.67$^{+0.06}_{-0.07}$ & 8.59$^{+0.24}_{-0.28}$ & PNLF & \cite{Herrmann2008} \\
29.98 $\pm$ 0.28  & 9.91 $\pm$ 1.28 & SN II standard candle & \cite{Olivares2010} \\
29.76 $\pm$ 0.24 & 8.95 $\pm$ 0.99 & SN II photospheric magnitude & \cite{Rodriguez2014} \\
29.74 $\pm$ 0.18 & 8.87 $\pm$ 0.74 & SN II photospheric magnitude & \cite{Rodriguez2014} \\
30.01 $\pm$ 0.07 & 10.05 $\pm$ 0.32 & SN II photospheric magnitude & \cite{Rodriguez2014} \\
29.88 $\pm$ 0.05 & 9.46 $\pm$ 0.22 & SN II photospheric magnitude & \cite{Rodriguez2014} \\
30.04 $\pm$ 0.03 & 10.19 $\pm$ 0.14 & TRGB & \cite{Jang2014}  \\
29.91$^{+0.08}_{-0.13}$ & 9.59$^{+0.35}_{-0.57}$ & PNLF & this work
  \enddata
\end{deluxetable*}

\subsection{An IFU approach to PN studies}

Instruments like MUSE, with high spatial resolution and large field of view, are well suited to identifying PNe populations within nearby galaxies, but there has been limited application of such an IFU approach so far \citep{Roth2004, Sarzi2011}.  We have shown that at a physical resolution of 50\, pc, this technique can recover the PNLF cleaned of contaminants to measure a robust distance to the galaxy.  
This is most important for galaxies with a substantial amount of cold ISM, where supernova ejecta shocks are likely to create [OIII] 5007 emission. In systems without a cold ISM (elliptical galaxies, spiral bulges), the ejecta expands more freely, and will not normally create [O III] emission.  
Given the large field covered by our observations, we explore at what spatial scale our simple method for selecting PNe outlined in this paper breaks down.  

At  100pc scales (2$\arcsec$ resolution) we can identify 36 
[OIII] sources with line ratios consistent with PNe.  In most cases both the H$\alpha$ and [SII] lines are not detected above the background diffuse emission, making removal of SNR contaminants difficult without more detailed background modeling.  We find 23 
of these objects are in our PNe catalog, and four 
are in our SNR catalog.  The remaining are typically within $\sim$5$\arcsec$ from a cataloged PN, and are likely PNe that are blended with a neighboring object,  shifting the centroid.  

At 150 pc scales (3$\arcsec$ resolution) we identify 24 
[OIII] sources with line ratios consistent with PNe, 13 
of which are in our PNe catalog and three 
of which are in the SNR catalog.

At scales larger than 150pc we see extensive blending of the [OIII] emission, as we are tracing mainly the crowded, central and highly star-forming regions.  Crowded-field 3D spectrophotometry \citep{Roth2004} can be used to remove background due to unresolved stars and diffuse ISM emission and recover faint emission lines for detailed PN modeling.  More involved source extraction techniques, including use of prior knowledge of source positions from other high resolution imaging \citep{Kamann2013}, would be necessary to extract useful PN line fluxes on scales larger than this.  

As the PN population traces the stellar density, there is a clear advantage to pursuing PN studies within galaxy disks.  With narrowband imaging these environments were challenging due to the high stellar and ISM background, however with full spectroscopic information the search for PNe in these regions can recover a large number of objects with relatively small spatial coverage.  
For our MUSE pointings we measure the stellar mass surface density based on the Spitzer imaging where emission from dust and other nonstellar sources has been removed \citep{Querejeta2015}, and we estimate the V-band surface brightness based on a published radial profile \citep{Fisher2008}.
Our southern pointing (at slightly larger radius) has an average stellar surface density of 6.9 $\times$ 10$^7$ M$_\sun$ kpc$^{-2}$ (22.5 mag arcsec$^{-2}$) and contains 1.6 PNe per kpc$^2$, while the inner pointings at roughly twice the stellar surface density (1.9 $\times$ 10$^8$ M$_\sun$ kpc$^{-2}$, 21.0 mag arcsec$^{-2}$) contain twice the PN density (3.1 PNe per kpc$^{2}$).   Counting only the PNe detected above our completeness limit and suitable for fitting the PNLF, we find on average 1.3 PNe per kpc$^2$ in the inner pointings.

For a galaxy twice as far as NGC 628 ($\sim$20 Mpc) the PNLF cutoff would be 1.5 mag fainter.  To robustly trace the PNLF cutoff the required PN survey depth is 0.8 mag, however even a small sample (20 PNe) still can accurately recover the distance to within 3\% with only $\sim$10\% increased uncertainty \citep{Jacoby1997}.   Given our observed PN density, this corresponds to a survey area of $\sim$15 kpc$^2$ within the central galaxy disk.  For our shallower observations (42 minutes on source) with 27.0 mag completeness limit, this means we could measure a distance out to D=13.6 Mpc (1\arcsec=80 pc).  Our deeper southern pointings (50 minutes on source), reaching 27.5 mag completeness limit,  could measure a distance out to D=17.1 Mpc (1\arcsec=80 pc).  In both cases, a single MUSE observation in Wide Field Mode would cover the required survey area at a spatial resolution ($<$100 pc) sufficient for clean separation of a significant number of PNe.  

Optical IFU identification of PNe has applications not just for measuring distances but also for using PNe to trace chemical enrichment, through detailed study of their metallicity (\citealt{Ali2016}, and references therein), and stellar disk dynamics, through study of the emission line kinematics (\citealt{cortesi2013}, and references therein).   

\section{Conclusion}
\label{sec:conclusion}
We use the optical IFU instrument MUSE on the VLT to map [OIII], H$\alpha$, [NII] and [SII] emission lines across 27 kpc$^2$ within the central star-forming galaxy disk of NGC 628.  We select unresolved [OIII] emission line sources, and using diagnostic line ratios we distinguish 87 HII regions and 30 SNRs (Table \ref{tab:snr}) as contaminants to our sample.  

In total we identify 63 PNe within our fields (Table \ref{tab:pns}), and using the 36 PNe with m$_{[{\rm OIII}]} < 27.0$ mag we determine a new PNLF distance modulus of 29.91$^{+0.08}_{-0.13}$ mag (9.59$^{+0.35}_{-0.57}$ Mpc).  This is in good agreement with literature values, particularly with the TRGB estimate (which has the smallest uncertainty).  Our measured distance is significantly larger  than the previously reported PNLF distance, however we can reproduce this lower distance estimate by biasing our sample with SNR contaminants, demonstrating the power of full spectral information over narrowband imaging when isolating the PN sample. 

Given our limited spatial coverage within the galaxy, we show that this technique can be used to refine distance estimates even when only limited IFU observations are available.  In particular, when targeting central regions where the stellar density is high, IFU techniques can recover large numbers of PNe ($\sim$3 PNe per kpc$^2$) within galaxies at 10-20 Mpc distances (50-100 pc at 1$\arcsec$ resolution) that were previously unobservable due to confusion with background stellar and ISM emission.

\acknowledgements
KK thanks Iskren Georgiev and Jay Gallagher for their helpful discussions.  
We would like to thank the referee for their careful reading and comments.  
Based on observations made with ESO Telescopes at the La Silla Paranal Observatory under programme ID 094.C-0623 and ID 095.C-0473.
KK acknowledges grant KR 4598/1-2.  
B.G. gratefully acknowledges the support of the Australian Research Council as the recipient of a Future Fellowship (FT140101202).  
G.B. is supported by CONICYT/FONDECYT, Programa de Iniciaci—n, Folio 11150220.  
JMDK gratefully acknowledges support in the form of an Emmy Noether Research Group from the Deutsche Forschungsgemeinschaft (DFG), grant number KR4801/1-1.

This work was carried out as part of the Star Formation and Feedback in Nearby Galaxies (SFNG) collaboration. 

Funding for the SDSS and SDSS-II has been provided by the Alfred P. Sloan Foundation, the Participating Institutions, the National Science Foundation, the U.S. Department of Energy, the National Aeronautics and Space Administration, the Japanese Monbukagakusho, the Max Planck Society, and the Higher Education Funding Council for England. The SDSS Web Site is http://www.sdss.org/.

The SDSS is managed by the Astrophysical Research Consortium for the Participating Institutions. The Participating Institutions are the American Museum of Natural History, Astrophysical Institute Potsdam, University of Basel, University of Cambridge, Case Western Reserve University, University of Chicago, Drexel University, Fermilab, the Institute for Advanced Study, the Japan Participation Group, Johns Hopkins University, the Joint Institute for Nuclear Astrophysics, the Kavli Institute for Particle Astrophysics and Cosmology, the Korean Scientist Group, the Chinese Academy of Sciences (LAMOST), Los Alamos National Laboratory, the Max-Planck-Institute for Astronomy (MPIA), the Max-Planck-Institute for Astrophysics (MPA), New Mexico State University, Ohio State University, University of Pittsburgh, University of Portsmouth, Princeton University, the United States Naval Observatory, and the University of Washington.

\end{document}